\documentclass[iop,apj]{emulateapj}
\usepackage{apjfonts}
\usepackage{hyperref}

\journalinfo{The Astrophysical Journal}
\slugcomment{Accepted 2016 February 16}

\shortauthors{CAMILO ET AL.}
\shorttitle{RADIO DISAPPEARANCE OF THE MAGNETAR XTE~J1810$-$197}

\begin{document}


\def\xte{XTE~J1810$-$197}
\def\chandra{{\em Chandra}}
\def\xmm{{\em XMM-Newton}}

\title{Radio disappearance of the magnetar XTE~J1810$-$197 and
continued X-ray timing}

\author{F.~Camilo\altaffilmark{1,2},
  S.~M.~Ransom\altaffilmark{3},
  J.~P.~Halpern\altaffilmark{1},
  J.~A.~J.~Alford\altaffilmark{1},
  I.~Cognard\altaffilmark{4,5},
  J.~E.~Reynolds\altaffilmark{6},
  S.~Johnston\altaffilmark{6},
  J.~Sarkissian\altaffilmark{7},
  and W.~van~Straten\altaffilmark{8}
}

\altaffiltext{1}{Columbia Astrophysics Laboratory, Columbia University,
  New York, NY 10027, USA}
\altaffiltext{2}{SKA South Africa, Pinelands, 7405, South Africa}
\altaffiltext{3}{National Radio Astronomy Observatory, Charlottesville,
  VA 22903, USA}
\altaffiltext{4}{Laboratoire de Physique et Chimie de l'Environnement
  et de l'Espace, LPC2E CNRS-Universit{\'e} d'Orl{\'e}ans, F-45071
  Orl{\'e}ans, France}
\altaffiltext{5}{Station de radioastronomie de Nan\c{c}ay,
  Observatoire de Paris, CNRS/INSU F-18330 Nan\c{c}ay, France}
\altaffiltext{6}{CSIRO Astronomy and Space Science, Australia
  Telescope National Facility, Epping, NSW 1710, Australia}
\altaffiltext{7}{CSIRO Parkes Observatory, Parkes, NSW 2870, Australia}
\altaffiltext{8}{Swinburne University of Technology, Hawthorn, VIC
  3122, Australia}

\begin{abstract}
We report on timing, flux density, and polarimetric observations
of the transient magnetar and 5.54\,s radio pulsar \xte\ using the
Green Bank, Nan\c{c}ay, and Parkes radio telescopes beginning in
early 2006, until its sudden disappearance as a radio source in
late 2008.  Repeated observations through 2016 have not
detected radio pulsations again.  The torque on the neutron star,
as inferred from its rotation frequency derivative $\dot \nu$,
decreased in an unsteady manner by a factor of 3 in the first year
of radio monitoring, until approximately mid-2007.  In contrast,
during its final year as a detectable radio source, the torque
decreased steadily by only 9\%.  The period-averaged flux density,
after decreasing by a factor of 20 during the first
10 months of radio monitoring, remained relatively steady
in the next 22 months, at an average of $0.7\pm0.3$\,mJy
at 1.4\,GHz, while still showing day-to-day fluctuations by factors
of a few. There is evidence that during this last phase of
radio activity the magnetar had a steep radio spectrum, in contrast
to earlier flat-spectrum behavior.  There was no secular decrease
that presaged its radio demise.  During this time the pulse profile
continued to display large variations, and polarimetry, including
of a new profile component, indicates that the magnetic geometry
remained consistent with that of earlier times.  We supplement these
results with X-ray timing of the pulsar from its outburst in 2003
up to 2014.  For the first 4 years, \xte\ experienced non-monotonic
excursions in frequency derivative by at least a factor of 8.  But
since 2007, its $\dot\nu$ has remained relatively stable near its
minimum observed value.  The only apparent event in the X-ray record
that is possibly contemporaneous with the radio shut-down is a
decrease of $\approx20\%$ in the hot-spot flux in 2008--2009, to a
stable, minimum value.  However, the permanence of the high-amplitude,
thermal X-ray pulse, even after the (unexplained) radio demise,
implies continuing magnetar activity.

\end{abstract}


\section{Introduction}\label{sec:intro} 

Magnetars are ultra-highly magnetized neutron stars (inferred surface
dipolar field strengths $B_s \approx 10^{14-15}$\,G) that display
hugely variable and sometimes very bright X-ray emission, powered
by their decaying fields \citep{dt92a}.  This process is reflected
in their extremely unsteady rotation.  Their magnetic fields cause
magnetars to spin down very rapidly, and all those known have long
periods, $2<P<12$\,s.

Among the 23 known magnetars
\citep{ok14}\footnote{\url{http://www.physics.mcgill.ca/~pulsar/magnetar/main.html}.},
four are known to be transient emitters of radio pulsations.  The
first to be so identified was the $P=5.54$\,s anomalous X-ray pulsar
(AXP) \xte, discovered in early 2003 following an X-ray outburst
\citep{ims+04}.  It is unclear when radio emission started, but
pulsations were not present in 1998, while a point source was visible
by early 2004 \citep{hgb+05}.  Radio pulsations were detected in
early 2006 \citep{crh+06}, with some properties that are markedly
different from those of ordinary rotation-powered pulsars, including
extremely variable flux densities and pulse profiles, and flat
spectra \citep[e.g.,][]{crp+07,ljk+08}.  The emission is also highly
linearly polarized, like that of several ordinary young radio pulsars
\citep{crj+07,ksj+07}.  To a great extent, these properties are
shared by all four radio magnetars identified so far
\citep{crhr07,crj+08,lbb+10,kjlb11,sj13}.

The radio emission from \xte\ arose following the X-ray outburst.
The X-ray flux then decayed exponentially and returned to pre-outburst
levels in 2007--2008 \citep{bpg+11}.  Here we show that \xte\ ceased
to emit detectable radio pulsations in late 2008, and present the
timing, flux density, polarimetric, and pulse profile behavior
during its last 20 months of radio activity.  We also present X-ray
timing measurements and fluxes through 2014.

\section{Radio Observations}\label{sec:obs} 

Our previously published radio studies of \xte\ (also known as
PSR~J1809$-$1943) are based on extensive data sets largely from
2006.  Here we present results based on observations done with the
Robert C.\ Byrd Green Bank Telescope (GBT), the Nan\c{c}ay
radio telescope (NRT), and the CSIRO Parkes telescope, mainly
through the end of 2008.  Table~\ref{tab:sens} summarizes
relevant parameters for all the radio observations presented in
this paper.

\begin{deluxetable}{lllll}
\tabletypesize{\footnotesize}
\tablewidth{0pt}
\tablecolumns{5}
\tablecaption{\label{tab:sens}Parameters and Sensitivities for Radio Observations of \xte}
\tablehead{
\colhead{}  & \colhead{NRT} & \colhead{PKS AFB} & 
\colhead{PKS DFB} & \colhead{GBT} }
\startdata
Center frequency (MHz) & 1398   & 1374              & 1369     & 1950   \\
Bandwidth (MHz)        &   64   &  288              &  256     &  600   \\
Gain, $G$ (K\,Jy$^{-1}$) & 1.55\tablenotemark{a} & 0.735\tablenotemark{b} & 0.735\tablenotemark{b} & 1.9\tablenotemark{c} \\
System temperature, $T_{\rm sys}$ (K) & 47\tablenotemark{d} & 44 & 44 & 28 \\
SEFD\tablenotemark{e} (Jy) & 30\tablenotemark{f} & 60\tablenotemark{g} & 60\tablenotemark{g} & 15\tablenotemark{h} \\
$\eta$ & 1.0 & $\sqrt{(\pi/2)}$\tablenotemark{i} & 1.0 & 1.2\tablenotemark{j} \\
SEFD$_{\rm eff}$\tablenotemark{k} (Jy) & 30 & 76 & 60 & 17
\enddata
\tablecomments{Observations used the
Berkeley-Orl\'eans-Nan\c{c}ay (BON) coherent dedispersor \citep{ct06} at the
NRT,
the analog \citep[AFB;][]{mlc+01} and digital
\citep[DFB;][]{mhb+13} filterbanks at Parkes, and the pulsar
Spigot \citep{kel+05} at the GBT. Sensitivity parameters are given in the
direction of \xte\ for the specified frequencies and bandwidths.
The sky temperature at 1.4\,GHz in this direction is $T_{\rm sky}
= 16.4$\,K including CMB (obtained from
\url{http://www3.mpifr-bonn.mpg.de/survey.html}; \citealt{rtr01}).}
\tablenotetext{a}{\citet{tch+05}.}
\tablenotetext{b}{Nominal $G$ of multibeam receiver center pixel
\citep{mlc+01}.}
\tablenotetext{c}{\url{http://www.gb.nrao.edu/~fghigo/gbtdoc/sens.html}.}
\tablenotetext{d}{$T_{\rm sky}$ plus ``no-sky'' $T_{\rm
sys} = 30$\,K (given cold-sky $T_{\rm sys} = 35\,$K;\\ \url{http://www.nrt.obspm.fr/nrt/obs/NRT_tech_info.html}).}
\tablenotetext{e}{System equivalent flux density ($\equiv T_{\rm sys}/G$).}
\tablenotetext{f}{Computed from $T_{\rm sys}$ and known $G$.}
\tablenotetext{g}{Calibrated assuming Hydra~A flux density of
42.5\,Jy at 1.4\,GHz \citep[after accounting for a 1.5\% beam
dilution factor;][]{bgpw77}.}
\tablenotetext{h}{Measured within Spigot band from flux-calibrated
GUPPI observation\\
(\url{https://safe.nrao.edu/wiki/bin/view/CICADA/GUPPiUsersGuide}).}
\tablenotetext{i}{Inefficiency factor due to AFB 1-bit sampling \citep{mlc+01}.}
\tablenotetext{j}{Estimated inefficiency factor due to Spigot 3-level
quantization \citep[cf.][]{kel+05}.}
\tablenotetext{k}{Effective SEFD ($\equiv \eta \times \mbox{SEFD}$),
used to compute \xte\ flux densities (see Sections~\ref{sec:flux}
and \ref{sec:spect}).}
\end{deluxetable}

\subsection{Radio Timing}\label{sec:timing}

In \citet{ccr+07} we showed the timing behavior of \xte\ through
2007 January, based largely on 1.4\,GHz data collected
with the BON spectrometer at the NRT.  As the flux density
decreased it became preferable to time the pulsar at the GBT.
We did this at 2\,GHz using the Spigot spectrometer,
recording the data in search mode and folding offline.
Nevertheless, BON timing continued to be important, particularly
during 2007 May--August when the GBT was not available. The
NRT observations were coherently dedispersed, with the full band
divided into 16 frequency channels, then folded at the predicted
pulsar period with 2 minute sub-integrations before mid-2007 and
30\,s thereafter. See Table~\ref{tab:toas} for a log of the timing
observations newly presented here.  Daily observations typically
lasted 1\,hr at Nan\c{c}ay and varied greatly at the GBT, from
0.25\,hr to 2\,hr with most at least 0.5\,hr (i.e., from a couple
of hundred to over 1000 pulsar rotations per session).

\begin{deluxetable}{lcc}
\tabletypesize{}
\tablewidth{0pt}
\tablecolumns{3}
\tablecaption{\label{tab:toas} Log of Radio Timing Observations of \xte}
\tablehead{
\colhead{MJD range (days)}  & \colhead{Number of daily TOAs} &
\colhead{Telescope} }
\startdata
54128--54218 (90)  & $53+18$ & $\mbox{GBT}+\mbox{Nan\c{c}ay}$ \\
54226--54357 (131) & 27    & Nan\c{c}ay                       \\
54352--54739 (387) & 51    & GBT
\enddata
\tablecomments{GBT TOAs were obtained with Spigot at 2\,GHz;
Nan\c{c}ay TOAs were obtained with BON at 1.4\,GHz (see
Section~\ref{sec:timing}). The average observing cadence in the
three periods listed decreased from once every 1.5 days, to once
every 5 days, to once every 7.5 days. }
\end{deluxetable}

In principle there are unusual challenges involved in radio timing
of \xte, because of changing pulse profiles.  In practice we found
that on the vast majority of days a simple procedure was sufficient
to obtain times-of-arrival (TOAs) that could be used to describe
reliably the rotation of the star.  We first excised strong
radio frequency interference from the BON and Spigot data. BON
timing was detailed in \citet{ccr+07}.  For Spigot we obtained TOAs
by cross-correlating individual folded pulse profiles with a Gaussian
template.  We then used the TOAs with the TEMPO
software\footnote{\url{http://tempo.sourceforge.net}.} to obtain
timing solutions.

\begin{figure}
\begin{center}
\includegraphics[angle=0,scale=0.42]{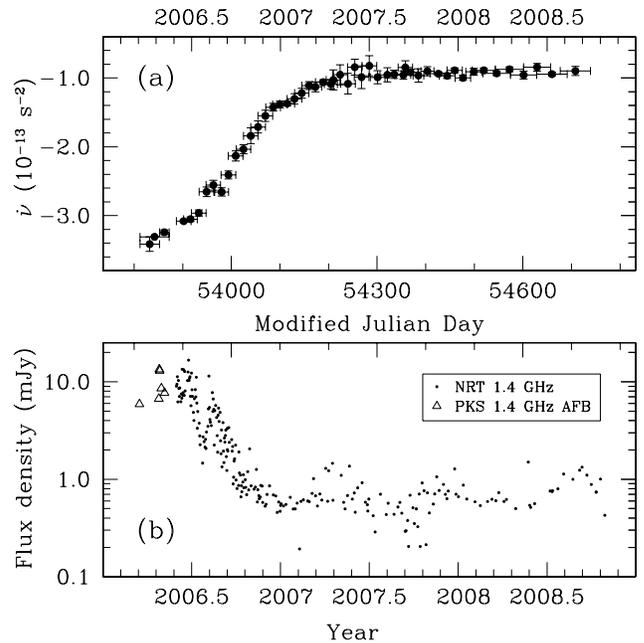}
\caption{\label{fig:fdotflux}
Torque and radio flux density of \xte.  (a) Frequency derivative
versus date (Section~\ref{sec:timing}).  Measurements from
before MJD~54127 (2007.1) are reproduced from \citet{ccr+07}. (b)
Period-averaged flux density versus date.  Data from before
2007.1 were presented in \citet{ccr+07} but have been reanalyzed
for this paper (see Section~\ref{sec:flux}).  }
\end{center}
\end{figure}

The position of \xte\ was held fixed in all our timing fits at that
determined from VLBI observations \citep{hcb+07}; the measured
proper motion is too small to affect the timing of this pulsar.  As
explained in \citet{ccr+07}, we maintained phase connection for
this pulsar since 2006 April, but the rotation frequency derivative
was changing so rapidly that it proved more informative to measure
$\dot \nu$ (where $\nu = 1/P$) by using TOAs typically spanning one
month and doing a TEMPO fit for only $\nu$ and $\dot \nu$.  We did
such fits in segments of data offset by roughly 15 days to provide
good sampling of $\dot \nu$, and the results are shown in
Figure~\ref{fig:fdotflux}a.  In this panel, the first 17 measurements
are reproduced from \citet{ccr+07}, while the next 33 measurements
are new.

\begin{figure}
\begin{center}
\includegraphics[angle=0,scale=0.42]{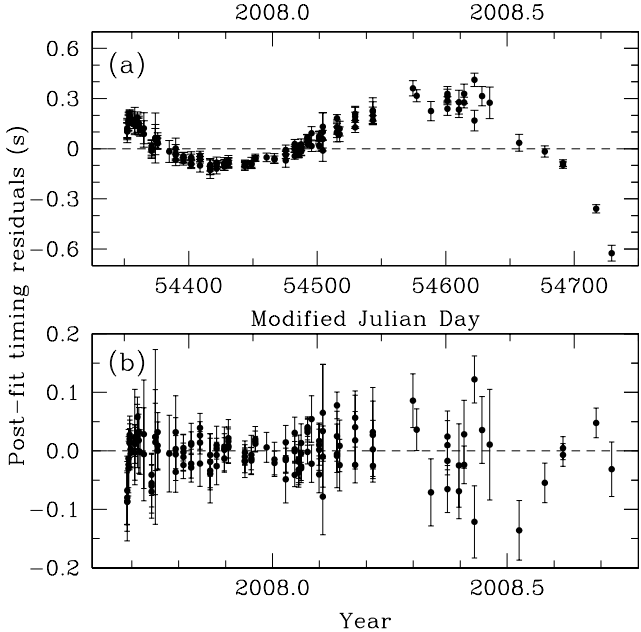}
\caption{\label{fig:timing}
Timing residuals for \xte.  (a) Phase residuals versus date for a
timing model that fits only for rotation phase, frequency, and
frequency derivative, showing a cubic trend.  (b) Residuals for a
model that fits for phase, $\nu$, $\dot \nu$, and $\ddot \nu$ (see
Section~\ref{sec:timing} and Table~\ref{tab:parms}).  }
\end{center}
\end{figure}

\begin{deluxetable}{ll}
\tabletypesize{}
\tablewidth{0pt}
\tablecolumns{2}
\tablecaption{\label{tab:parms} Two Radio Timing Solutions for \xte}
\tablehead{
\colhead{Parameter}  & \colhead{Value} }
\startdata
R.A. (J2000.0)                              & $18^{\rm h}09^{\rm m}51\fs087$  \\
Decl. (J2000.0)                             & $-19\arcdeg43\arcmin51\farcs93$ \\
Dispersion measure, DM                      & 178.0\,pc\,cm$^{-3}$            \\
Epoch (MJD TDB)                             & 54550.0                         \\
Range of dates (MJD)                        & 54352--54729                    \\
Frequency, $\nu$~\tablenotemark{a}          & 0.18048830377(8)\,Hz            \\
Frequency derivative, $\dot\nu$~\tablenotemark{a} & $-9.163(1)\times10^{-14}$\,Hz\,s$^{-1}$ \\
\tableline
Frequency, $\nu$~\tablenotemark{b}          & 0.1804882977(1)\,Hz    \\
Frequency derivative, $\dot\nu$~\tablenotemark{b} & $-9.090(2)\times10^{-14}$\,Hz\,s$^{-1}$ \\
Frequency second derivative, $\ddot\nu$~\tablenotemark{b} & $2.46(5)\times10^{-22}$\,Hz\,s$^{-2}$ \\
RMS post-fit timing residual ($P$)          & 0.007
\enddata
\tablecomments{The celestial coordinates were held fixed at the
values obtained from VLBA observations \citep{hcb+07}, and the DM
was held fixed at the value obtained from simultaneous 0.7 and
2.9\,GHz observations \citep{crh+06}. }
\tablenotetext{a}{These two parameters are sufficient to obtain a
phase-connected solution encompassing the MJD range, but do not
fully describe the rotation of the neutron star.  See
Figure~\ref{fig:timing}a and Section~\ref{sec:timing}. }
\tablenotetext{b}{These three parameters fully describe the rotation
of the neutron star within the given MJD range, but have little
predictive value outside it.  See Figure~\ref{fig:timing}b and
Section~\ref{sec:timing}. }
\end{deluxetable}

It is clear from Figure~\ref{fig:fdotflux}a that over the last $\sim
400$ days $\dot \nu$ stabilized greatly, compared to earlier
large variations.  To investigate this in detail we obtained
phase-connected fits spanning the last year of timing data.  In
Figure~\ref{fig:timing}a we show the residuals from a simple fit
to rotation phase, $\nu$, and $\dot \nu$, showing large residuals.
In Figure~\ref{fig:timing}b, we see that the addition of $\ddot
\nu$ absorbs all remaining residual trends.  These timing solutions
are listed in Table~\ref{tab:parms}.  The positive value of $\ddot
\nu$ implies that during this span $-\dot \nu$ (proportional to the
braking torque) was decreasing steadily, by a total of 9\% during
the year.

\subsection{Radio Flux Densities}\label{sec:flux}

One unusual aspect of radio emission from \xte\ is the fluctuation
on $\sim$ daily timescales of its period-averaged flux density,
largely intrinsic to the pulsar \citep[e.g.,][]{ljk+08}.  These
variations result from a combination of different pulse profile
components becoming active (i.e., because of radically changing
profiles) and varying intensity from particular components.
Superimposed on this apparently chaotic variation, the average flux
density of the pulsar decreased by over an order of magnitude in
the 10 months following its pulsed radio discovery, as
seen in Figure~\ref{fig:fdotflux}b (these pre-2007 data
were originally presented in \citealt{ccr+07} but have been reanalyzed
here in order to present a consistent flux density record).

Most period-averaged flux densities presented in this paper
(including all in Figure~\ref{fig:fdotflux}b) were obtained by
measuring for each daily observation the area under the pulse
profile, scaled to its off-pulse rms, and converting to a Jansky
scale using the observing parameters and telescope system
noise (cf.\ the SEFD$_{\rm eff}$ values in
Table~\ref{tab:sens}; see section 7.3.2 of \citealt{lk04} for more
details on this method).  We estimate that the absolute 1.4\,GHz
flux density scale is accurate to within 10\%.

The greatest source of uncertainty arises from the impact of radio
frequency interference (RFI) and system noise fluctuations on this
long period and, from 2007, faint pulsar.  Each pulse
profile was carefully excised of RFI in both the frequency and time
domains. Nevertheless, for approximately one-third of all
the post-2006 NRT observations, the RFI was so bad or the flux
density was so low (approaching the $\approx 0.1$\,mJy detection
threshold), that we did not extract a flux density measurement at
all. In some of the remaining instances, as much as half of the
data had to be discarded in order to obtain an integrated profile
clean enough to measure flux density. Mindful of these caveats (most
often some residual RFI is bound to remain, and individual measurements
could be greatly affected), we estimate that the NRT flux density
measurements in Figure~\ref{fig:fdotflux}b have a typical relative
fractional uncertainty of $\approx 20$\%, with a minimum of 0.1\,mJy.

Two things immediately stand out from Figure~\ref{fig:fdotflux}b:
daily flux density variations continued through the end of our data
set, by factors of a few; and the average 1.4\,GHz flux
density stabilized early in 2007, at a level of $0.68\pm0.27$\,mJy
for all post-2006 NRT detections presented here.  Also, the panels
of Figure~\ref{fig:fdotflux} are curiously something of a mirror
image to each other.

\subsubsection{Radio Disappearance}\label{sec:dis}

\begin{figure}
\begin{center}
\includegraphics[scale=0.90]{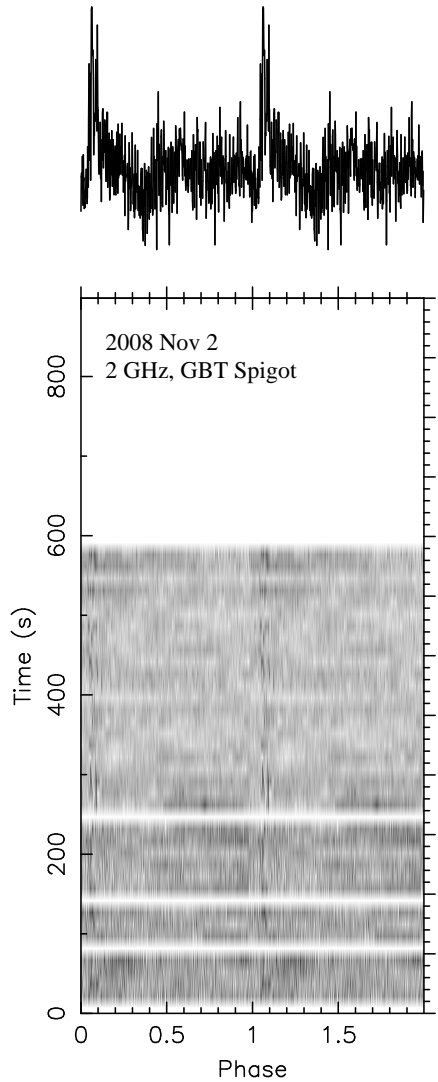}
\caption{\label{fig:lastGBT}
Last radio detection of \xte\ at the GBT. The period-averaged flux
density is among the smallest we observed ($\sim 50\,\mu$Jy),
but pulse detection is aided by the ``spiky'' nature of its narrow
sub-pulses.  Two rotations are shown as a function of time, with
the summed profile at the top. White areas are subintegrations
masked due to particularly bad RFI.  }
\end{center}
\end{figure}

\begin{figure}
\begin{center}
\includegraphics[angle=0,scale=0.42]{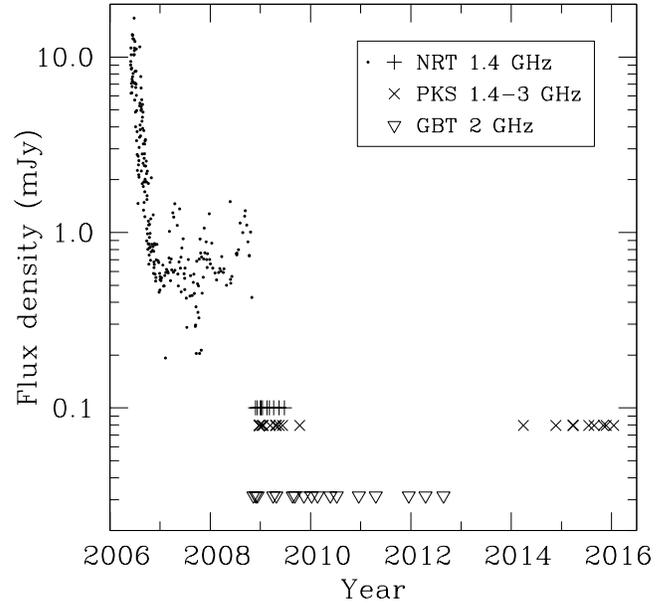}
\caption{\label{fig:nd}
Fourty-seven radio non-detections of \xte\ since late 2008
(Section~\ref{sec:dis}).  The dots represent measured
NRT flux densities, reproduced from Figure~\ref{fig:fdotflux}b.
The other symbols denote individual observations without a
detection, placed at approximately the upper flux density limits
for the respective sets of observations. }
\end{center}
\end{figure}

With no warning from either its timing or flux density behavior,
\xte\ ceased to emit detectable radio pulsations in late 2008.  The
last detection at the NRT was on October 29, on November
2 at the GBT (Figure~\ref{fig:lastGBT}), and at Parkes on the
following day.  The next attempt to detect it was on November 10.
We attempted to detect the pulsar at Parkes on 20 occasions
through 2016 January, largely at 1.4\,GHz, each time for 0.5--1\,hr.
At the NRT we did 10 more observations through 2009 June.
At the GBT we made a total of 17 attempts at 2\,GHz, each $\approx
0.5$\,hr, through 2012 August.  The pulsar was not detected in any
of these 47 observations spanning 7 years since
2008 November (Figure~\ref{fig:nd}).  We emphasize that \xte\ did
not gradually fade into undetectability.  A few weeks before the
last detection, we recorded beautiful profiles (e.g.,
Figures~\ref{fig:pol}d, \ref{fig:profs}c and \ref{fig:profs}d); the
signal strength and pulse profiles were fluctuating at least as
much as they had for the previous $\sim 500$ days.  And then the
radio pulses were gone.

\begin{figure}
\begin{center}
\includegraphics[angle=0,scale=0.42]{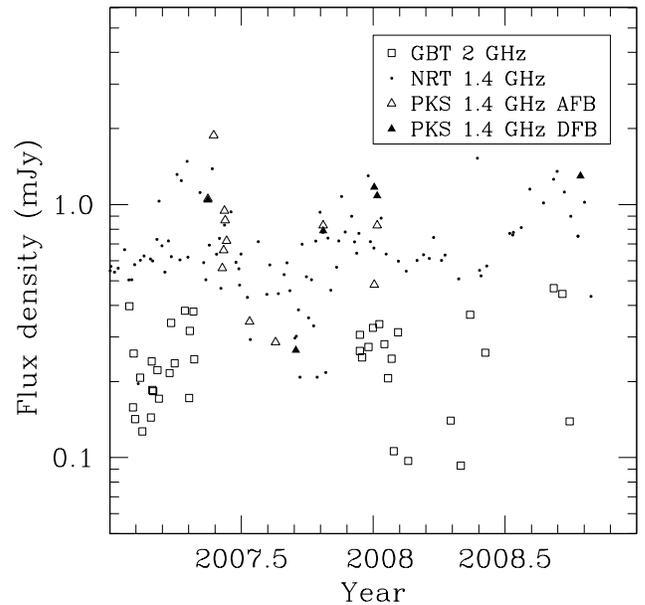}
\caption{\label{fig:alpha}
Flux density measurements for \xte\ at 1.4\,GHz and 2\,GHz
in 2007--2008.  Parkes observations include those done with analog
(AFB) and full-Stokes digital (DFB) filterbanks.  See
Section~\ref{sec:spect} for details. }
\end{center}
\end{figure}

\begin{figure}
\begin{center}
\includegraphics[angle=0,scale=1.06]{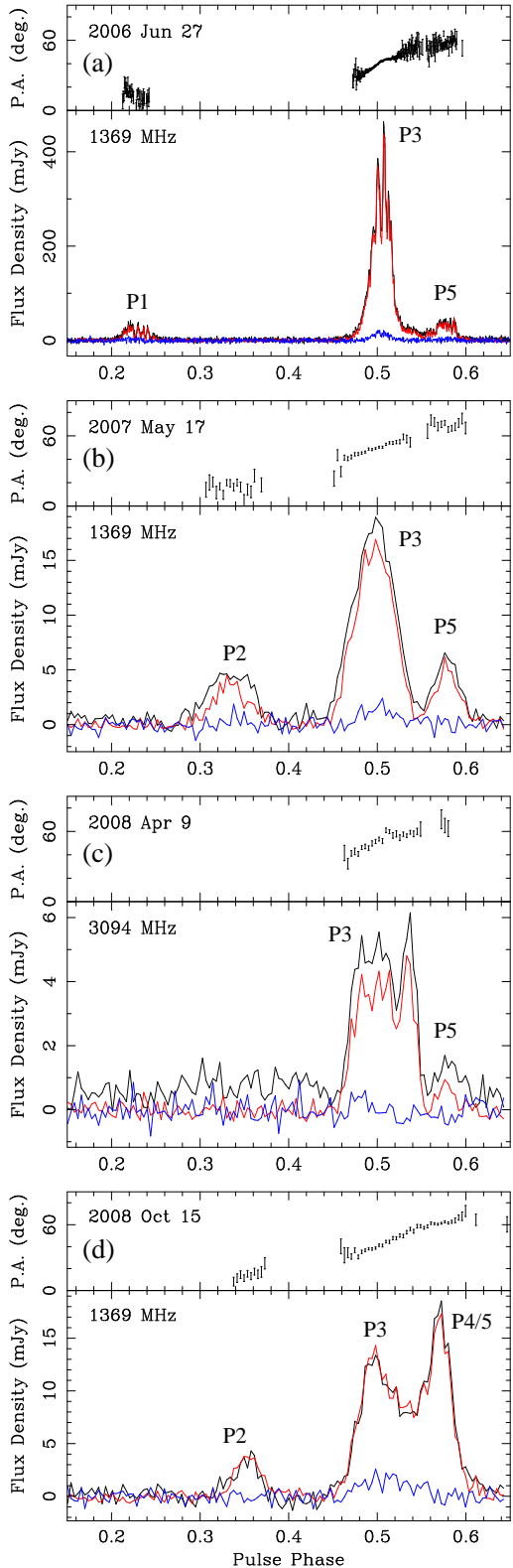}
\caption{\label{fig:pol}
Polarimetric pulse profiles of \xte\ recorded at Parkes.  Only 50\%
of pulse phase is shown. In each lower panel the black trace
represents total intensity, red is linear polarization, and blue
is circular.  Each upper panel shows the position angle of linear
polarization, rotated to the pulsar frame using
$\mbox{RM}=78$\,rad\,m$^{-2}$.  The profiles are phase-aligned with
respect to each other by eye, and absolute phase is arbitrary.
Recurring profile features are labelled P1--P5 (as in
Figure~\ref{fig:profs}). }
\end{center}
\end{figure}

For an assumed pulse duty cycle of 6\% (comparable to the
component widths often observed for this pulsar) and the parameters
used in our monitoring observations (Table~\ref{tab:sens}), the
upper limit on the flux density of \xte\ since late 2008 is
approximately 0.1\,mJy at 1.4\,GHz based on the NRT observations,
slightly lower than that for the Parkes observations, and approximately
0.03\,mJy at 2\,GHz based on the GBT observations (Figure~\ref{fig:nd}).
These limits are nearly an order of magnitude below the {\em
average\/} flux densities during the 1.8 years of radio emission
post-2006, although not much below the faintest established detections
(see Figures~\ref{fig:lastGBT} and \ref{fig:alpha}).  The great
profile variability, long period, and RFI, make more detailed
estimates unreliable. In any case, since the last detection in
late 2008, in none of 47 observations spanning 7
years was the pulsar as detectable as in the poorest of more than
200 detections made over a period of 2 years
before 2008 November.

\subsection{Radio Spectrum}\label{sec:spect}

The available evidence suggests that during the faint epoch
that lasted for 2 years preceding its radio disappearance in late
2008, \xte\ had a steep radio spectrum, contrasting to its earlier
generally flat spectrum.

In Figure~\ref{fig:alpha} we present all our post-2006 flux
density measurements at 1.4\,GHz and 2\,GHz. The 1.4\,GHz NRT values
are reproduced from Figure~\ref{fig:fdotflux}b. These are fundamentally
consistent with Parkes measurements at the same frequency from six
full-Stokes observations using pulsar digital filterbanks (DFBs),
analyzed with PSRCHIVE \citep{hvm04}, and 12 analog filterbank (AFB)
observations.

The 2\,GHz flux density values presented here were obtained
from a subset of the GBT data used to derive TOAs
(Section~\ref{sec:timing}), using the method outlined in
Section~\ref{sec:flux} for NRT and Parkes AFB observations.  We
determined the system noise (Table~\ref{tab:sens}) from a full-Stokes
2\,GHz observation in the direction of \xte, flux-calibrated with
PSRCHIVE.  Even after careful RFI excision, we selected for reliable
flux density measurements only 40\% of all observations from which
we extracted a TOA.  We estimate relative fractional uncertainties
of $\approx 20$\% on average.

The 2\,GHz measurements in Figure~\ref{fig:alpha} range
over 0.1--0.5\,mJy, with an average and standard deviation of $S_2
= 0.25\pm0.10$\,mJy.  This is to be compared to $S_{1.4} =
0.68\pm0.27$\,mJy for the 1.4\,GHz measurements in the figure
(Section~\ref{sec:flux}), spanning much the same time interval. At
face value this would seem to suggest a spectral index of $\alpha
\approx -3$ (where $S_\nu \propto \nu^{\alpha}$).  Given the inherent
difficulties in extracting such a measurement for this variable
pulsar from non-simultaneous multi-frequency observations susceptible
to RFI despite our best efforts, we do not claim a reliable numerical
value for $\alpha$.  But Figure~\ref{fig:alpha} strongly suggests
that \xte\ indeed became a steep-spectrum object during its final
``weak'' state prior to disappearance as a radio source, in that
respect more akin to an ordinary pulsar than earlier in its ``high''
state, when the torque was also varying rapidly
(Figure~\ref{fig:fdotflux}).

\subsection{Polarimetry}\label{sec:pol}

All of our previously published \xte\ polarimetric data are from
before 2006 December \citep{crj+07}.  The data published by
\citet{ksj+07} end even earlier, but include single-pulse polarimetry.
Here we present some polarimetric observations done in 2007 and
2008 with Parkes at 1.4\,GHz and 3\,GHz, using respectively the
center beam of the multibeam receiver and the 10\,cm band of the
1050cm receiver.  The data were collected using DFBs, and
analyzed with PSRCHIVE, as in \citet{crj+07}.

Figure~\ref{fig:pol}a is a reprocessed version of a 2006 observation
presented in \citet{crj+07}, differing mainly in the flux
density scale and amount of data excised due to RFI.  The recalculated
rotation measure is $\mbox{RM} = 76\pm1$\,rad\,m$^{-2}$, entirely
consistent with the value in \citet{crj+07}, and the RMs calculated
for the latter observations are consistent with this value within
their larger uncertainties.

The 3\,GHz full-Stokes profile (Figure~\ref{fig:pol}c) looks similar
to its counterpart from 1.5 years before \citep[Figure~1c of][]{crj+07}.
At 1.4\,GHz, the two 2007/2008 profiles (Figure~\ref{fig:pol}b and
\ref{fig:pol}d) show very similar position angles of linear
polarization (PA), and differ mainly in the relative amplitudes of
three total-intensity profile components (labelled P2, P3, and
P4/5). Both profiles are close to 100\% linearly polarized.

Comparison between the 2006 profile (Figure~\ref{fig:pol}a) and the
2007/2008 1.4\,GHz profiles shows that the PA sweep and its absolute
values are similar for the ``main pulse'' regions (components
P3--P5).  However, the 2007/2008 profiles show component P2 not
present in the 2006 profile --- whose P1 component in turn is not
seen later on.  Profile components P1--P5 altogether span 40\% of
pulse phase, and if they all display stable PAs between 2006 and
2007/2008, then in principle this allows for a more conclusive
investigation of geometry than previously possible (see
Section~\ref{sec:rvm}).

\subsection{Radio Pulse Profiles}\label{sec:profs}

\begin{figure}
\begin{center}
\includegraphics[angle=0,scale=1.21]{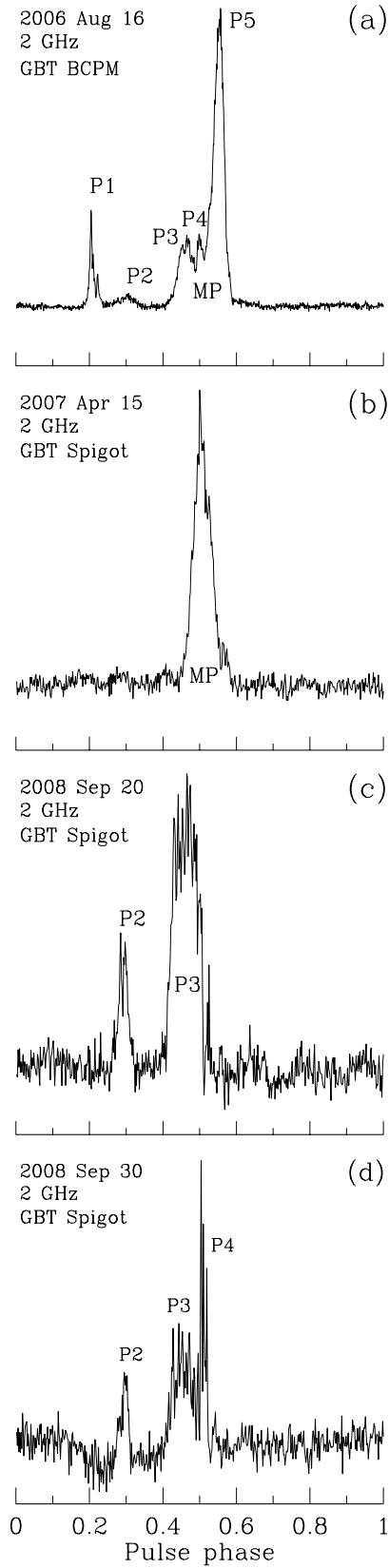}
\caption{\footnotesize \label{fig:profs}
A selection of \xte\ radio pulse profiles. Prominent features in
the profile that reoccur are labelled P1--P5 (as in Figure~\ref{fig:pol}).
We define the ``main pulse'' (MP) region as being composed of up
to features P3--P5.  Profiles are aligned by eye, and the absolute
phase here is arbitrary.  }
\end{center}
\end{figure}

The total-intensity pulse profiles of \xte\ were always extremely
variable, both in phase of active emitting regions and in the daily
appearance of each pulse profile component.  The description that
follows is based on a review of hundreds of daily profiles that we
obtained during 2006--2008 at the GBT, Parkes, and Nan\c{c}ay.
There was never any stable pulse profile, but we detected emission
from some pulse longitudes more often than from others, with some
discernible patterns over time, and radio pulsations were never
detected from many longitudes.  In that sense, there was some
long-lasting stability to the radio-emitting region.

Referring to Figure~\ref{fig:profs}a, emission was always detected
from the main pulse region (MP, in turn made up of at least three
discernible sub-structures, labelled P3--P5, not all necessarily
emitting at once). Emission from component P1 (see also
Figure~\ref{fig:pol}a) was detected only during 2006, and it was
the most variable, sometimes being much brighter than the MP
components.  On the other hand, P2 was active sometimes in 2006 and
2007 (Figure~\ref{fig:pol}b), but became more common in 2008
(Figures~\ref{fig:pol}d and \ref{fig:profs}c--d).  In 2007 the
observed profile often consisted of emission from the MP region
alone (Figure~\ref{fig:profs}b). In only one (2006) observation out
of hundreds did we detect emission from all these regions at once
(Figure~\ref{fig:profs}a).

The relative amplitudes of the different components varied widely,
exemplified in Figure~\ref{fig:profs}d by P4, usually not preeminent
but on this day the brightest of any.  The very ``spiky'' nature
of the individual sub-pulses that built up the broader integrated
profile components \citep[see, e.g.,][]{ssw+09} continued to the
very end (Figure~\ref{fig:lastGBT}).

Our sense is that in 2007--2008 the profiles were less variable
than in 2006, but this impression may be biased by two factors: the
pulsar was much brighter in 2006, which allowed the detection of
very faint rarely observed components \citep[see Figure~2 of][]{ccr+07};
and the total integration (both in number and in average duration
of observations) was larger in 2006 compared to later.  In any case
it is clear that during the period when the torque had stabilized
by comparison to earlier huge variations, and when the period-averaged
flux density had also stabilized in an average sense at a low level
(Figure~\ref{fig:fdotflux}), the pulse profiles of \xte\ were still
varying at unprecedented levels compared to normal pulsars, and
continued to do so until radio pulsations disappeared.

\section{X-ray Observations}\label{sec:xray}

In order to search for clues to the disappearance of radio pulsations
from \xte, we reviewed all its archival \chandra\ and \xmm\ data
collected from 2003--2014.  Timing results from 2009 November onward
are newly published here.  Detailed spectroscopic and flux analysis
are presented in \citet{ah16}, with a summary of the fluxes
given in Section~\ref{sec:xrayflux}.  In summary, we find that the
X-ray fluxes stepped down to a minimum value around the time that
the radio pulsations shut off, and this is the only recognizable
event in X-rays that is plausibly contemporaneous with the disappearance
in radio.

\subsection{X-ray Timing}\label{sec:xraytiming}

\begin{deluxetable*}{lrlcrl}
\tabletypesize{\small}
\tablewidth{0pt}
\tablecolumns{6}
\tablecaption{Log of X-ray Timing Observations of \xte}
\tablehead{\colhead{Mission/Instrument} & \colhead{ObsID } & \colhead{Date} &
\colhead{Epoch} & \colhead{\hfill Exposure} & \colhead{Frequency} \\
& & \colhead{(UT)} & \colhead{(MJD)} & \colhead{(ks)} &  \colhead{(Hz)}}
\startdata
\chandra\ HRC    & 4454       & 2003 Aug 27 & 52878 &  2.8 & 0.180531(11)    \\
\xmm\ pn+MOS     & 0161360301 & 2003 Sep  8 & 52890 & 12.1 & 0.18052682(30)  \\
\xmm\ pn         & 0152833201 & 2003 Oct 12 & 52924 &  8.9 & 0.1805245(12)   \\
\chandra\ HRC    & 5240       & 2003 Nov  1 & 52944 &  2.8 & 0.180536(10)    \\
\xmm\ pn+MOS     & 0161360501 & 2004 Mar 11 & 53075 & 18.9 & 0.18052415(25)  \\
\xmm\ pn+MOS     & 0164560601 & 2004 Sep 18 & 53266 & 28.9 & 0.18051856(18)  \\
\xmm\ pn+MOS     & 0301270501 & 2005 Mar 18 & 53447 & 42.2 & 0.18051142(16)  \\
\xmm\ pn+MOS     & 0301270401 & 2005 Sep 20 & 53633 & 42.2 & 0.1805046(3)    \\
\xmm\ pn+MOS     & 0301270301 & 2006 Mar 12 & 53806 & 51.4 & 0.1804991(4)    \\
\chandra\ ACIS-S & 6660       & 2006 Sep 10 & 53988 & 30.1 & 0.1804942(14)   \\
\xmm\ pn+MOS     & 0406800601 & 2006 Sep 24 & 54002 & 50.3 & 0.18049355(34)  \\
\xmm\ pn+MOS     & 0406800701 & 2007 Mar  6 & 54165 & 68.3 & 0.18049117(27)  \\
\xmm\ pn+MOS     & 0504650201 & 2007 Sep 16 & 54359 & 74.9 & 0.18048987(19)  \\
\chandra\ ACIS-S & 7594       & 2008 Mar 18 & 54543 & 29.6 & 0.1804868(15)   \\
\xmm\ pn+MOS     & 0552800201 & 2009 Mar  5 & 54895 & 65.8 & 0.18048610(24)  \\
\xmm\ pn+MOS     & 0605990201 & 2009 Sep  5 & 55079 & 21.6 & 0.1804857(13)   \\
\xmm\ pn+MOS     & 0605990301 & 2009 Sep  7 & 55081 & 19.9 & 0.1804829(16)   \\
\xmm\ pn+MOS     & 0605990401 & 2009 Sep 23 & 55097 & 14.2 & 0.1804872(24)   \\
\chandra\ ACIS-S & 11102      & 2009 Nov  1 & 55136 & 25.1 & 0.1804852(16)   \\
\chandra\ ACIS-S & 12105      & 2010 Feb 15 & 55242 & 12.6 & 0.180476(6)     \\
\chandra\ ACIS-S & 11103      & 2010 Feb 17 & 55244 & 12.6 & 0.180484(6)     \\
\xmm\ pn+MOS     & 0605990501 & 2010 Apr  9 & 55295 &  9.9 & 0.1804820(48)   \\
\chandra\ ACIS-S & 12221      & 2010 Jun 07 & 55354 & 10.0 & 0.180489(8)     \\
\xmm\ pn+MOS     & 0605990601 & 2010 Sep  5 & 55444 & 11.3 & 0.1804771(40)   \\
\chandra\ ACIS-S & 13149      & 2010 Oct 25 & 55494 & 15.4 & 0.1804766(35)   \\
\chandra\ ACIS-S & 13217      & 2011 Feb  8 & 55600 & 15.0 & 0.1804835(35)   \\
\xmm\ pn+MOS     & 0671060101 & 2011 Apr  3 & 55654 & 22.9 & 0.1804796(13)   \\
\xmm\ pn+MOS     & 0671060201 & 2011 Sep  9 & 55813 & 15.9 & 0.1804751(18)   \\
\chandra\ ACIS-S & 13746      & 2012 Feb 19 & 55976 & 20.0 & 0.1804790(23)   \\
\chandra\ ACIS-S & 13747      & 2012 May 24 & 56071 & 20.0 & 0.1804774(21)   \\
\xmm\ pn+MOS     & 0691070301 & 2012 Sep  6 & 56176 & 17.9 & 0.1804721(17)   \\
\xmm\ pn+MOS     & 0691070401 & 2013 Mar  3 & 56354 & 17.9 & 0.1804722(18)   \\
\xmm\ pn+MOS     & 0720780201 & 2013 Sep  5 & 56540 & 24.5 & 0.1804721(12)   \\
\chandra\ ACIS-S & 15870      & 2014 Mar  1 & 56717 & 20.1 & 0.1804723(20)   \\
\xmm\ pn+MOS     & 0720780301 & 2014 Mar  4 & 56720 & 25.0 & 0.1804682(11)   \\
\chandra\ ACIS-S & 15871      & 2014 Sep  7 & 56911 & 20.1 & 0.1804710(23)
\enddata
\label{tab:xraylog}
\end{deluxetable*}

\begin{figure}
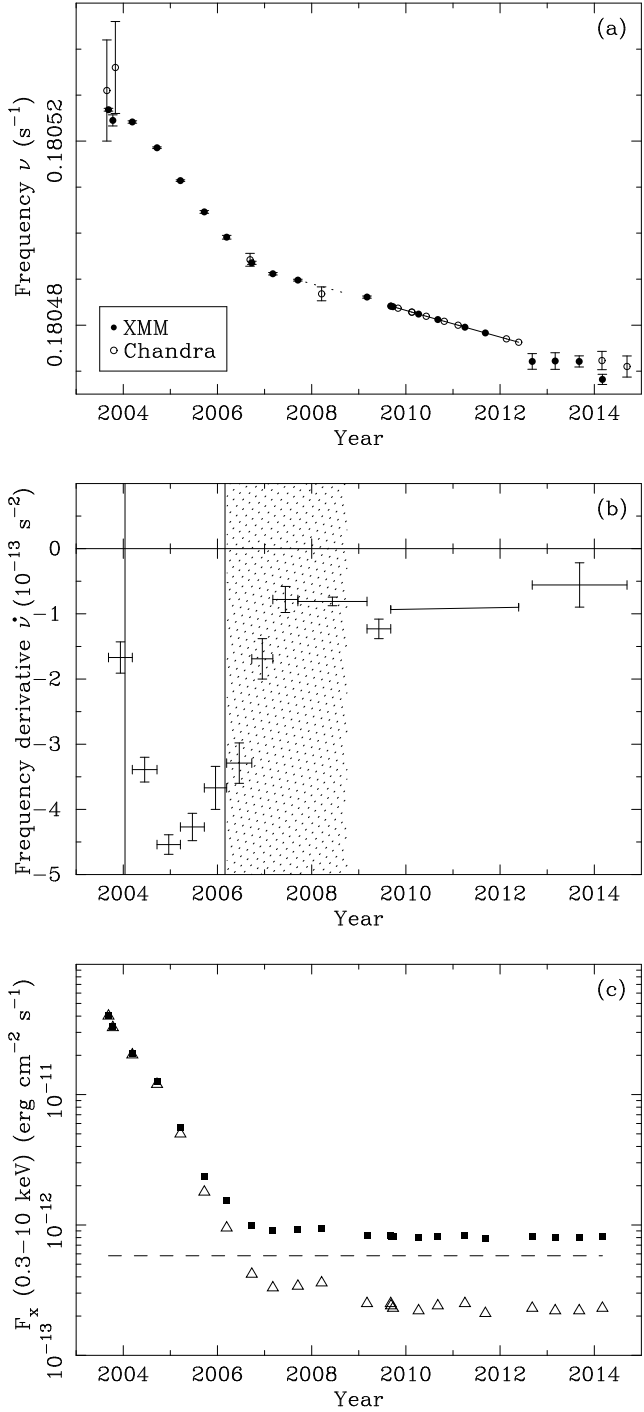

\begin{center}
\includegraphics[width=0.98\linewidth]{fig8a.eps}
\end{center}
\vspace{-0.1in}
\begin{center}
\includegraphics[width=0.98\linewidth]{fig8b.eps}
\end{center}
\vspace{-0.1in}
\begin{center}
\includegraphics[width=0.98\linewidth]{fig8c.eps}
\end{center}
\vspace{-0.1in}
\caption{\footnotesize \label{fig:xray}
X-ray timing and flux properties of \xte\ from \chandra\ and \xmm.
(a) Frequency measurements, where the points linked with
the solid line in 2009--2012 comprise the phase-connected solution
in Table~\ref{tab:xraytime}.  The dashed line segment is the radio
timing solution from Table~\ref{tab:parms}.  (b) Frequency
derivatives obtained by differencing adjacent frequency measurements,
and from the 2009--2012 phase-connected solution.  The vertical
lines denote the first two epochs of radio detection at 1.4\,GHz
with the VLA \citep{hgb+05,crh+06}, and the shaded region encompasses
the epochs of pulsed radio detection from Figure~\ref{fig:fdotflux}.
It is not known whether radio emission at the flux density level
of the first detection ($4.5\pm 0.5$\,mJy) was present earlier in
the X-ray outburst, which was detected in 2003 January.  (c)
X-ray flux measurements from the three- or two-temperature blackbody
fits of \citet{ah16}.  Open triangles are the sum of the varying
hot and warm areas, and filled squares are the total flux including
the triangles and the cooler full surface area of the neutron star
represented by the (constant) dashed line. }
\end{figure}

Table~\ref{tab:xraylog} is a log of all \xte\ timing observations
performed by \chandra\ and \xmm\ through 2014.  The \chandra\ ACIS
observations were taken with the source on the S3 CCD, and with a
subarray of 100 or 128 rows to obtain time resolution of 0.3\,s or
0.4\,s, respectively. \xmm\ observations used the pn CCD in full-frame
or large-window mode, with 74\,ms or 44\,ms resolution and, in most
cases, the MOS CCDs in small-window mode with 0.3\,s resolution.
We corrected the processed archival \xmm\ photons for leap seconds
and time jumps when needed, and applied the barycentric correction
at the VLBI measured position.  We then computed frequencies and
$1\,\sigma$ errors using the $Z_1^2$ test on photons in the 0.3--4\,keV
band.  The two short \chandra\ HRC observations that were obtained
for the purpose of source location have relatively uncertain
frequencies, and were not used in the subsequent analysis.

The frequency measurements are shown in Figure~\ref{fig:xray}a.
These were used to track the time-varying frequency derivative from
2003--2009 by computing the difference in frequency between adjacent
\xmm\ observations, usually 6 months apart. Where these
$\dot \nu$ measurements overlap with the more precise record obtained
from radio observations during 2006--2008 (Figure~\ref{fig:fdotflux}a),
they agree. Figure~\ref{fig:xray}b illustrates that $\dot \nu$
varied by a factor of $\approx 6$.  In the months immediately
following the outburst, which was first detected in 2003 January,
monitoring by {\em RXTE\/} showed noisy spin-down with an even
higher mean frequency derivative of $-6.7\times 10^{-13}$\,Hz\,s$^{-1}$
\citep{ims+04}, which is $\approx 8$ times the minimum value of
$\approx -8 \times 10^{-14}$\,Hz\,s$^{-1}$ measured by \xmm\ in
2007--2008 (Figure~\ref{fig:xray}b).

Starting in 2009 September, more frequent observations allowed a
phase-connected solution to be established up through 2012 May.
This process was facilitated by the relatively stable spin-down
with small rate at later times.  Beginning with the two observations
on 2009 September 5 and 7, we folded the photons jointly using the
$Z_1^2$ test, and used the resulting frequency to fold the next
observation, verifying that the predicted phase agrees with the
observed one to $<0.2$ cycles.  Each subsequent observation was
then added to the joint fit and the  $Z_1^2$ test was iterated with
free parameters $\nu$ and $\dot \nu$, and finally $\ddot \nu$, until
15 observations were included.  The resulting fit is shown by the
solid line in Figure~\ref{fig:xray}a, and its $\dot \nu = -9.21
\times 10^{-14}$\,Hz\,s$^{-1}$ (Table~\ref{tab:xraytime}) is very
nearly a continuation of the minimum spindown rate that was first
reached in 2007.  The second frequency derivative as listed in
Table~\ref{tab:xraytime} is necessary to fit the 2009--2012 series
with a continuous ephemeris.

\begin{deluxetable}{ll}
\tabletypesize{}
\tablewidth{0pt}
\tablecolumns{2}
\tablecaption{X-ray Timing Solution for \xte}
\tablehead{
\colhead{Parameter} & \colhead{Value}
}
\startdata
R.A. (J2000.0)                   & $18^{\rm h}09^{\rm m}51\fs087$             \\
Decl. (J2000.0)                  & $-19\arcdeg43\arcmin51\farcs93$            \\
Epoch (MJD TDB)                  & 55444.0                                    \\
Range of dates (MJD)             & 55079--56071                               \\
Frequency, $\nu$                 & 0.18048121539(63)\,Hz                      \\
Frequency derivative, $\dot \nu$ & $-9.2121(35) \times 10^{-14}$\,Hz\,s$^{-1}$\\
Frequency second derivative, $\ddot \nu$ & $4.1(3)\times10^{-23}$\,Hz\,s$^{-2}$ \\
Surface dipole magnetic field, $B_s$\tablenotemark{a} & $1.3\times 10^{14}$\,G\\
Spin-down luminosity, $\dot E$\tablenotemark{b}       & $6.6\times 10^{32}$\,erg\,s$^{-1}$ \\
Characteristic age, $\tau_c$\tablenotemark{c}         & 31\,kyr 
\enddata
\tablenotetext{a}{$B_s = 3.2\times10^{19} (P\dot P)^{1/2}$\,G, with
$P$ in s, where $P=1/\nu$.}
\tablenotetext{b}{$\dot E = 4 \pi^2 \times 10^{45} \dot
P/P^3$\,erg\,s$^{-1}$.}
\tablenotetext{c}{$\tau_c = P/(2\dot P)$.}
\label{tab:xraytime}
\end{deluxetable}

Analyzing the same data from which we established the phase-connected
ephemeris of Table~\ref{tab:xraytime}, \citet{pbm+16} claimed to
identify a timing anomaly in which the pulsar was spinning up
(positive $\dot\nu$) between 2010 September and 2011 February.  This
could have arisen from their underestimation of uncertainties in
frequency measurements. In any case, our phase-coherent ephemeris
spanning this time shows no such event. \citet{pbm+16} also propose
that there is a single phase-connected timing solution spanning all
of 2007--2014. This is clearly invalid, because their listed
uncertainties on the polynomial coefficients ($\nu$, $\dot \nu$,
$\ddot \nu$) are two orders of magnitude larger than what would be
needed to describe a unique cycle count. Also, the parameters of
our actual phase-coherent timing segments in radio (Table~\ref{tab:parms})
and X-ray (Table~\ref{tab:xraytime}) disagree with theirs.  In
particular, their fitted $\dot\nu=-4.9\times10^{-14}$\,Hz\,s$^{-1}$
is about half of the true value.

After 2012 May we were unable to maintain phase connection.  The
prior coherent timing solution fails to predict the phase of the
2012 September observation by $\approx0.5$ cycles.  Possibly an
anti-glitch \citep{akn+13,sag14} and/or change in torque occurred
between 2012 May and September.  But we cannot tell for sure what
happened because the change in frequency differs from the extrapolation
by only $2\,\sigma$.  We can only estimate the frequency derivative
from 2012 September to 2014 September with an incoherent fit to the
frequencies, which is highly uncertain.  All we can say is that the
long-term frequency derivative did not clearly change in 2012 (see
Figure~\ref{fig:xray}b).

From 2009 September to 2012 May, $\dot \nu$ changed by at most 4\%,
as measured by the frequency second derivative (which is a factor
of 6 smaller than that measured over the year following 2007
September; Table~\ref{tab:parms}).  Such behavior is common in AXPs
and soft gamma repeaters (SGRs), which often show extended periods
of smooth spin-down at small $|\dot \nu|$ as well as noisier epochs
with larger $|\dot \nu|$
\citep{kcs99,kgc+01,gk02,wkg+02,wkf+07,tgd+08,dkg09,dk14}.  At one
time, it was proposed that the combined action of free and radiative
precession could explain ``bumpy'' spin-down of AXPs \citep{mel99,mel00},
which predicted that $\dot \nu$ would oscillate with a period of
several years.  However, as observations of dramatic changes in
$\dot \nu$ from magnetars have accumulated over the years, none are
apparently dominated by such periodic variations (see
references above). There is at best some evidence possibly
indicating quasi-periodic $\dot \nu$ behavior in one magnetar
\citep{akn+15}. The timing behavior of \xte\ shown in
Figure~\ref{fig:xray} seems typical when compared with other magnetars
that have been monitored for longer times with no cyclic pattern
evident in their spin-down.

\begin{figure*}
\begin{center}
\includegraphics[width=0.90\linewidth]{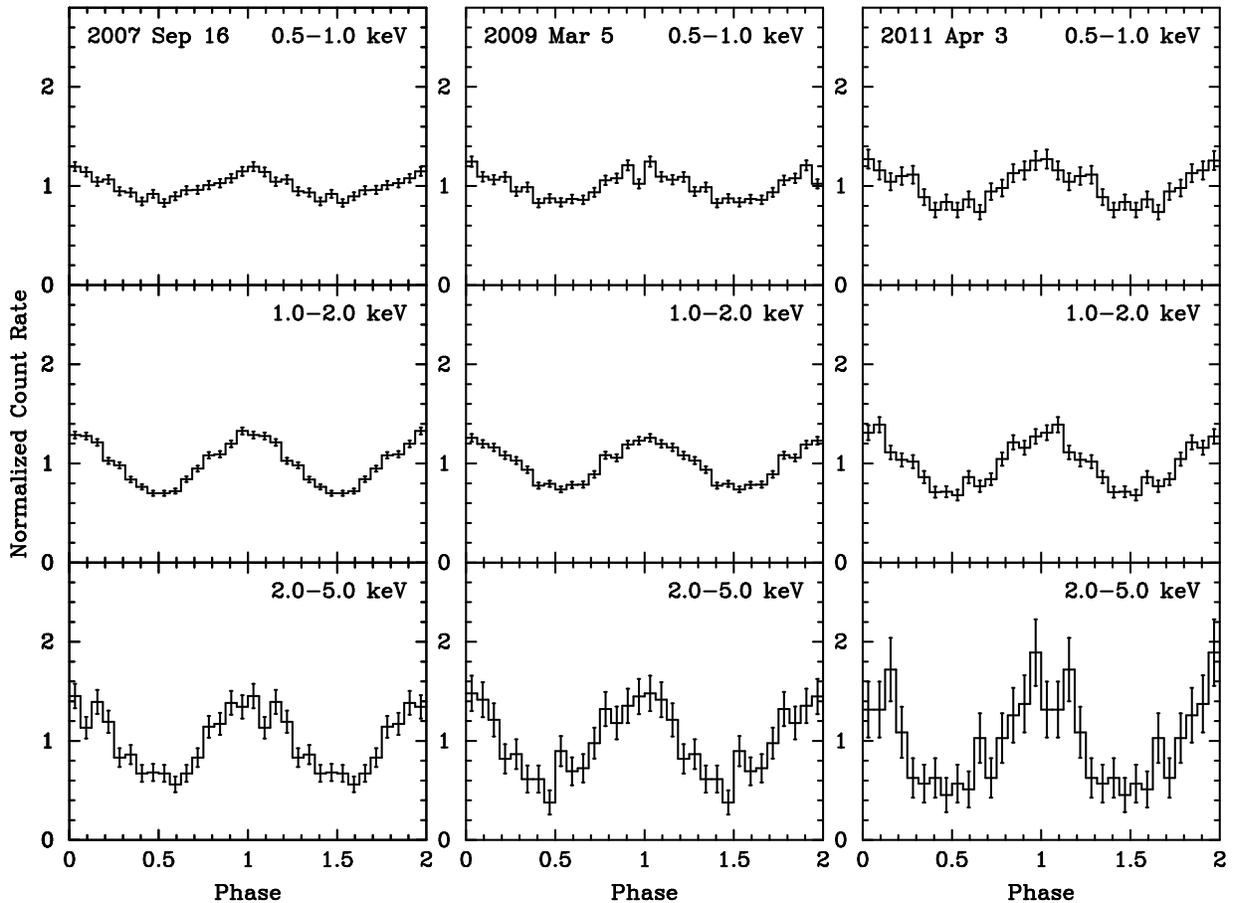}
\end{center}
\caption{\label{fig:xraypulse}
Energy-dependent pulse profiles of \xte\ from three \xmm\ observations.
Background has been subtracted, and the counts per bin are normalized
so that the average is 1 in each panel. The profiles are centered
on phase 1, which is arbitrary with respect to the radio phases as
displayed in Figures~\ref{fig:pol} and \ref{fig:profs}. }
\end{figure*}

Figure~\ref{fig:xraypulse} shows three sample energy-dependent X-ray
pulse profiles from \xmm, during (2007) and after (2009, 2011) the
epoch of pulsed radio detection. They share the same characteristics
as earlier observations, in particular those from 2005 and 2006 as
shown in \citet{gh07b}.  The pulse peaks are in phase as a function
of energy, while the pulsed fraction increases sharply with energy.
This is understood in a model in which a small hot spot is surrounded
by a cooler, larger annulus that covers most of the neutron star.
In summary, there is no obvious change in X-ray timing or pulse
shapes corresponding to the shut-down of radio emission in late
2008.

\subsection{X-ray Fluxes}\label{sec:xrayflux}

Figure~\ref{fig:xray}c summarizes the results from blackbody spectral
modelling that is described in more detail in \citet{ah16}.  The
spectrum is fitted with either two or three blackbodies, where the
coolest one is restricted to having a constant temperature and area
representing the full surface of the neutron star (dashed line in
Figure~\ref{fig:xray}c).  The triangles represent the fluxes from
one or two blackbodies (hot and warm) of much smaller area that
account for the decaying outburst flux.  The squares are the total
flux at each epoch.  It is evident that the fluxes have been steady
since after 2009, just after a slight decrease of $\approx20\%$ in
the hot/warm component between 2008 March and 2009 March.  This
small step-down in flux, corresponding to $\sim8\times10^{32}$
erg~s$^{-1}$ in bolometric luminosity, is the only event in X-rays
that we have been able to find that is approximately coincident
with the turn-off of radio pulsations in late 2008.

\subsection{X-ray and Radio Pulse Phase Alignment}\label{sec:xrayradio}

Five of the X-ray observations listed in Table~\ref{tab:xraylog}
were made during the period in which \xte\ was an active radio
source (see also Figure~\ref{fig:xray}).  For these observations,
we can therefore compare the emission phases of the X-ray and radio
profiles.  For each of these observations (two made with \chandra\
and three with \xmm), we used TEMPO to fit for an offset between
the X-ray TOA and a small number of surrounding radio TOAs (the
latter corrected to infinite frequency using the DM from
Table~\ref{tab:parms}).  The radio reference phase is the peak of
the main radio component, which owing to changing pulse shapes can
contribute up to $0.02\,P$ jitter in this comparison.  The X-ray
reference phase is the peak of the approximately sinusoidal profile,
which has an effective uncertainty of up to $0.1\,P$ for the \chandra\
observations and is a little better for the \xmm\ profiles.

The measured offset between the X-ray and the radio TOAs was, in
chronological order, 0.21, 0.03, $-0.17$, 0.01, and $-0.28$\,s.
All radio and X-ray TOAs therefore match to within $0.05\,P$.  Thus,
at least between 2006 September and 2008 March, the peak of radio
emission coincided with the peak of X-ray emission.

\section{Discussion}\label{sec:disc}

\subsection{Geometry of the Rotating Neutron Star}\label{sec:rvm}

Since the main radio pulse coincides with the peak of the X-ray
emission, it is reasonable to assume that both radio and X-ray are
coming from near the surface, where a hot spot is located at the
footpoint of a bundle of magnetic field lines \citep[for
the other two magnetars with detected radio and X-ray pulsations,
the radio profile is not as well aligned with, but still overlaps,
the X-ray profile;][]{hgr+08,tpe+15}.  In this picture,
currents flowing along this field-line bundle are responsible for
both surface heating and radio emission.  The X-ray and radio pulses
should coincide as long as the field lines are normal to the surface
and the radio emission height is much smaller than the radius of
the speed of light cylinder.  While these are reasonable assumptions
for the open field-line bundle in ordinary pulsars, the geometry
may be different in the case of magnetars, where currents on closed,
twisted magnetic field lines are thought to contribute to X-ray
emission via resonant cyclotron scattering \citep{lg06,ft07}, and
the location of the radio emission region is not obvious.  Therefore,
the rotating vector model \citep[RVM;][]{rc69a} is not necessarily
applicable unless the radio emission is produced on open field lines
as in ordinary pulsars.  Theory and observation are inconclusive
as to whether radio emission in magnetars is produced on open or
closed magnetic field lines \citep{tho08a}.

Assuming that the RVM applies to magnetars, \citet{crj+07} used it
to deduce the geometry of the star.  There were two possible solutions
depending on whether an orthogonal jump in polarization was introduced
between widely spaced pulse components (P1 and P3 in
Figure~\ref{fig:pol}a).  In the first solution, without an orthogonal
jump, the magnetic and rotational axes are almost aligned,
$\alpha\approx4\degr$ and $\beta\approx4\degr$, where $\alpha$ is
the angle between the magnetic and rotation axes, and $\beta$ is
the angle of closest approach of the line-of-sight to the magnetic
axis.  The second solution, with an orthogonal jump inserted, had
$\alpha\approx70\degr$ and $\beta\approx25\degr$.

New data reported here show a polarized component not present earlier
(P2 in Figure~\ref{fig:pol}b and d, at pulse phase between those
of P1 and P3). If we combine all the information on the PA values
from different epochs (namely from data represented in
Figures~\ref{fig:pol}a, b, and d), we obtain RVM fits which are
consistent with those presented in \citet{crj+07}. The result is
again $\alpha\approx4\degr$, $\beta\approx4\degr$ with no orthogonal
jump.  The addition of an orthogonal jump between components P2 and
P3 (see Figure~\ref{fig:pol}b) yields $\alpha\approx67\degr\pm10\degr$
and $\beta\approx16\degr\pm5\degr$.  Broadly speaking, therefore,
the same arguments apply as before --- either the pulsar is aligned,
which is problematic given the X-ray pulse properties discussed
below, or $\alpha$ is large and the emission height, determined
from the width of the pulse, is also relatively large,
$\sim2\times10^4$\,km, about 8\% of the light cylinder radius
\citep[see][]{crj+07}.

It appears that the X-ray spectrum of \xte\ has always been dominated
by the thermal hot spot and surrounding warm region \citep{gh05,hg05},
which is supported by the single-peaked X-ray pulse that aligns
well in phase as a function of energy.  So the X-ray pulse has been
modeled independently of the radio as an indicator of the spin
orientation and viewing geometry of a surface hot spot, which can
then be compared with the results of the RVM fits under the assumption
that the hot spot underlies a perpendicular radio beam.  The most
recent such X-ray modeling results \citep{bpg+11} allow, in our
notation, $\alpha$ in the range $29^{\circ}$--$52^{\circ}$, which
is degenerate with $\zeta=|\beta+\alpha|$, while $\beta$ ranges
from $0^{\circ}$ to $23^{\circ}$.  Their extreme solutions have
$(\alpha,\zeta)=(29^{\circ},52^{\circ})$ or
$(\alpha,\zeta)=(52^{\circ},29^{\circ})$.  Neither of the RVM
solutions are entirely consistent with the model of the X-ray pulse,
although the high pulsed fraction of the harder X-rays would at
least seem to rule out a nearly aligned rotator with $\alpha\sim4\degr$,
and a geometry with $(\alpha,\beta)\approx(52\degr,16\degr)$ seems
reasonably compatible with both radio and X-ray observations.

\subsection{Decline of the X-ray and Radio Luminosity}\label{sec:decline}

A detailed explanation for the exponential X-ray decay of \xte\ was
developed by \citet{bel09}.  In this model, a bundle of closed,
twisted magnetic field lines centered on the magnetic dipole axis
carries the current that heats a spot on the surface.  This so-called
``j-bundle'' is the result of a twist of the crust by the starquake
that initiated the outburst.  As the j-bundle untwists, its boundary
recedes toward the magnetic pole; thus, the area of its footpoint
decreases, which accounts for the declining blackbody area fitted
to the X-ray spectrum \citep{gh07b,bid+09}.

\citet{bel09} also accounts for the non-monotonic change in spin-down
rate of \xte, which after the outburst first increased then decreased
as shown in Figure~\ref{fig:xray}b.  The initial increase in torque
is caused by growing twist of field lines near the magnetic axis,
even as the outer boundary of the j-bundle is shrinking.  This twist
inflates the poloidal field lines, effectively increasing the dipole
moment and the magnetic field strength at the light cylinder.  Once
the twist reaches a maximum stable value of $\sim 1$ radian, the
dipole moment decreases due to the continued contracting of the
j-bundle, and the torque decreases.

In this picture, radio emission is produced on the closed field
lines of the j-bundle, which is wider and more energetic than the
open field-line bundle.  Therefore, radio emission from a transient
magnetar in outburst could be easier to see, both for geometric and
energetic reasons, than in its quiescent state.  Its radio beam
could be broader and different in many respects (spectrum, polarization,
variability) from those of ordinary radio pulsars.  The radio pulse
in this model is coincident with the X-ray pulse, and when the
j-bundle contracts to less than the width of the observed radio
beam, then the radio emission should decrease rapidly.

However, in this model one may expect the width of the radio pulse
to decrease gradually to zero as the j-bundle shrinks, possibly
approaching zero width and disappearing when the outer boundary of
the emitting region reaches the tangent point to the observer's
line of sight.  But this is contrary to the observations, which
show no change in the width or complexity of the radio pulse just
before it disappeared.  The X-ray flux from the hot/warm spot showed
at most a 20\% decrease ($\sim8\times10^{32}$ erg~s$^{-1}$) at the
epoch of radio disappearance, when there was no detectable change
in spin-down power in the X-ray timing.  Since this luminosity
change is comparable to or greater than the spin-down power at the
time, it is difficult to understand how spin power could be responsible
for the event.  Probably a more subtle physical process is required
to explain the sudden quenching of the radio emission.

\citet{rptt12} proposed that the (then) three radio-detected magnetars
share the property that their quiescent X-ray luminosity is smaller
than their spin-down power, but not the converse: not all magnetars
with $L_x/\dot E < 1$ are radio pulsars.  However, review of these
parameters for \xte\ shows that the minimum spin-down power of \xte,
which it reached in quiescence in 2007--2012, is in the range $\dot
E = (5.6-6.6) \times 10^{32}$\,erg\,s$^{-1}$, while its quiescent
(0.3--10\,keV) X-ray luminosity is $L_x \approx 1 \times
10^{33}\,d_{3.5}^2$\,erg\,s$^{-1}$, uncorrected for absorption,
both before and after the outburst \citep{ghbb04,bpg+11,ah16}.  The
distance of 3.5\,kpc is taken from \citet{mcr+08}.  The bolometric
luminosity of the cool component alone from \citet{ah16} is
$4\times10^{34}\,d_{3.5}^2$\,erg\,s$^{-1}$.  These values of $\dot
E$ and $L_x$ disagree with the ones used in \citet{rptt12}, and do
not support their proposition, because the X-ray luminosity of \xte\
is greater than its spin-down power, whether before, during, or
after the outburst.

\citet{smg15} explain radio pulsations from magnetars and ordinary
radio pulsars by a single model, the partially screened gap.  It
assumes that rotational energy heats the open-field-line polar cap,
and the resulting temperature, compared with a critical temperature
for ion emission, is what determines whether a partially screened
gap is maintained.  Only if the luminosity of the polar cap is much
less than the spin-down power is radio emission possible.  However,
the temperature of the X-ray/radio emitting cap in \xte\ is larger
than that of a rotation-powered pulsar with the same timing parameters
because it is heated by magnetic field decay, not by rotation.
During the outburst of \xte\ the polar cap luminosity rose by more
than two orders of magnitude while the spin-down luminosity only
increased by a factor of 8.  So it is not clear how this model could
explain the onset or turn-off of radio pulsations during the outburst
of \xte.

\section{Conclusions}

Radio pulsations appear to be characteristic of some, but not all
transient magnetars in outburst.  The long record left by the single
known outburst of \xte\ provides a prototype for investigating the
mechanism of magnetar radio emission.  The radio flux densities and
X-ray fluxes each declined by about a factor of 50 from
the peak of the outburst to the year 2006.  Then the X-ray and radio
luminosities both levelled off in 2007.  This large-amplitude
correlation would argue that the power for the radio emission comes
from the magnetar mechanism that creates currents in the pulsar
magnetosphere and heats the neutron star crust, rather than from
rotation power.

The radio pulses of \xte\ before they turned off continued to show
large day-to-day fluctuations, unlike ordinary radio pulsars.
The radio spectrum appears to have changed from flat to
steep as the radio (and X-ray) emission leveled off.  The emission
remained highly polarized, and the observation of a new polarized
pulse component allowed us to test and refine the previously derived
emission geometry assuming a dipole field geometry.  This was
compared with independent modeling of the X-ray pulse, which is
coincident in phase with the main radio pulse.  The radio polarization
allows two solutions, depending on whether an orthogonal jump in
polarization is assumed between pulse components.  However, the
almost aligned solution appears inconsistent with the large-amplitude
X-ray pulse, so we favor the more inclined model.

Finally, the radio pulsations turned off abruptly in late 2008, and
have not reappeared in the subsequent 7 years.  However, a continuing
pulse of hard X-rays from a hot spot persists during the radio quiet
epoch and exceeds the spin-down luminosity, evidence of continuing
magnetar activity.   This should not be thought of as a ``return
to quiescence,'' because magnetar activity is not a quiescent state,
but a continuing conversion of magnetic energy to luminosity that,
for a period of years, may well result in a quasi-constant luminosity.
Also, we do not know enough about the pre-outburst state of \xte\
to determine if it had precisely the same emission properties then
as is does now, or if it was truly quiescent prior to the outburst
detected in early 2003.

The sudden radio disappearance prompted us to search for any
contemporaneous event in the X-ray record that could be associated.
The only possible such occurrence was a step-down in the X-ray flux
of the hot spot by $\approx20\%$ between 2008 March and 2009 March.
Although this would appear to be a small effect, it bears some
consideration, as none of the other observations and theories offer
a natural explanation for the sharp radio turn-off.

\acknowledgements

The National Radio Astronomy Observatory is a facility of the
National Science Foundation operated under cooperative agreement
by Associated Universities, Inc.

The Nan\c{c}ay Radio Observatory is operated by the Paris Observatory,
associated with the French Centre National de la Recherche Scientifique
(CNRS).

The Parkes Observatory is part of the Australia Telescope, which
is funded by the Commonwealth of Australia for operation as a
National Facility managed by CSIRO.

This investigation is partly based on observations obtained with
\xmm, an ESA science mission with instruments and contributions
directly funded by ESA Member States and NASA, and with \chandra.
The \chandra\ X-ray Observatory Center is operated by the Smithsonian
Astrophysical Observatory for and on behalf of NASA under contract
NAS8-03060.  This research also made use of data obtained from the
High Energy Astrophysics Science Archive Research Center (HEASARC),
provided by NASA's Goddard Space Flight Center. JPH acknowledges
support from NASA ADAP grant NNX15AE63G.

{\em Facilities:}  \facility{GBT (Spigot, BCPM)}, \facility{NRT
(BON)}, \facility{Parkes (DFB, AFB)}, \facility{CXO (ACIS-S)},
\facility{XMM (pn, MOS)}

\end{document}